\documentclass[%
 reprint,
 amsmath,amssymb,
 aps,
]{revtex4-2}

\usepackage{graphicx}
\usepackage{dcolumn}
\usepackage{bm}
\usepackage{color}
\usepackage[normalem]{ulem}

\usepackage[mathlines]{lineno}
\usepackage{comment}

\begin{document}

\preprint{APS/123-QED}

\title{Divergence-free algorithms for solving nonlinear differential equations on quantum computers}

\author{Katsuhiro Endo}
 \email{katsuhiro.endo@aist.go.jp}
\author{Kazuaki Z. Takahashi}%
 \email{kazu.takahashi@aist.go.jp}
\affiliation{%
 National Institute of Advanced Industrial Science and Technology (AIST),\\
 Research Center for Computational Design of Advanced Functional Materials,\\
 Central 2, 1-1-1 Umezono, Tsukuba, Ibaraki 305-8568, Japan
}%

\date{\today}

\begin{abstract}
From weather to neural networks, modeling is not only useful for understanding various phenomena, but also has a wide range of potential applications. Although nonlinear differential equations are extremely useful tools in modeling, their solutions are difficult to obtain. Based on the expectation of quantum transcendence, quantum algorithms for efficiently solving nonlinear differential equations continue to be developed. However, even the latest promising algorithms have been pointed out to have an evolution time limit. This limit is the theoretically predestined divergence of solutions. We propose algorithms of divergence-free simulation for nonlinear differential equations in quantum computers. For Hamiltonian simulations, a pivot state $\bm{s}$ in the neighborhood of state $\bm{x}$ is introduced. Divergence of the solutions is prevented by moving $\bm{s}$ to a neighborhood of $\bm{x}$ whenever $\bm{x}$ leaves the neighborhood of $\bm{s}$. Since updating $\bm{s}$ is directly related to computational cost, to minimize the number of updates, the nonlinear differential equations are approximated by nonlinear polynomials around $\bm{s}$, which are then Carleman linearized. Hamiltonian simulations of nonlinear differential equations based on several representative models are performed to show that the proposed method breaks through the theoretical evolution time limit. The solution of nonlinear differential equations free from evolution time constraints opens the door to practical applications of quantum computers.
\end{abstract}

\maketitle


\begin{figure*}
\includegraphics[width=1\textwidth]{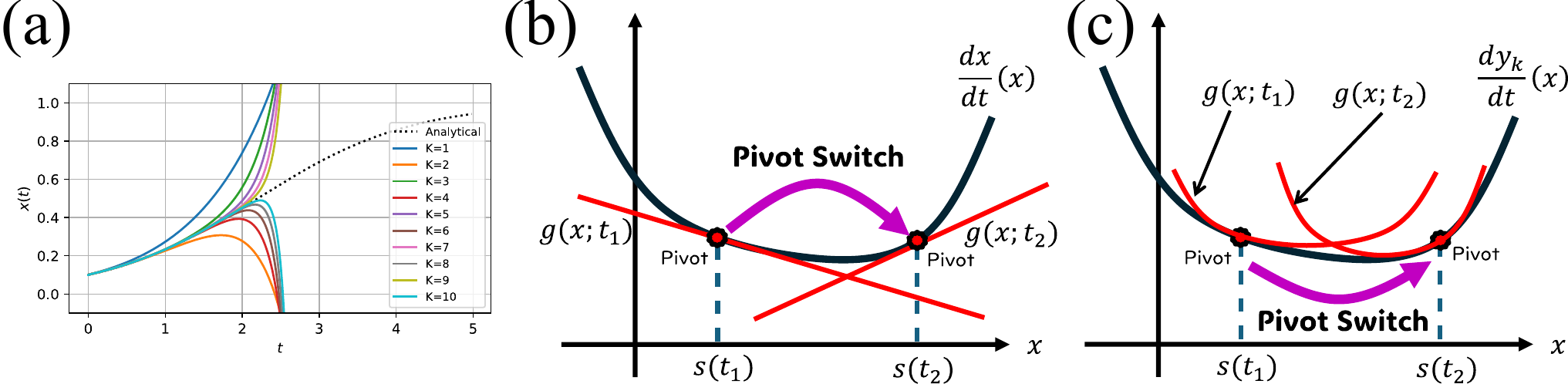}
\caption{(a) Time evolution of solutions of the conventional Carleman linearization-based algorithm with various $K$ for the logistic equation. The dashed line is analytical solution. All simulations using the algorithm diverge regardless of the value of $K$.
(b and c) Conceptual illustrations of the PS and PSC methods proposed in this paper, respectively. $g(x,t)$ is a linear or polynomial approximation of the time derivative in the neighborhood of the pivot state $s$.
}
\label{fig:1}
\end{figure*}

\section{\label{sec:intro}Introduction}
Modeling, the attempt to create a well-characterized model to better capture a subset of the world, is an essential and inseparable part of many scientific disciplines, including physics, biology, medicine, engineering, and economics \cite{zeigler2000theory}. Good modeling based on deep insights can predict the future or evolve into fundamental technologies: Climate change models predict current global warming, and neural network models support machine learning technologies. Differential equations play a pivotal role in modeling. Nonlinear differential equations can represent in mathematical form even complex phenomena such as collective motion of molecules \cite{frenkel2023understanding}, atmospheric turbulence \cite{ferziger2019computational}, and the spread of infectious diseases \cite{hethcote2000mathematics}, where interactions affect themselves. However, obtaining solutions to nonlinear differential equations is much more difficult than formulating the equations. Generally, it is not easy to obtain an explicit formula for the general solution, and in most cases, numerical solutions are obtained by relying on computers. Moreover, if the behavior of the solutions changes significantly with slight differences in the initial conditions, long-term prediction is next to impossible, no matter how powerful classical computers are used.

Quantum computers are expected to have quantum transcendence, i.e., to perform certain calculations much more efficiently than their classical counterparts. Quantum transcendence has already been theoretically confirmed for unitary linear differential equation systems. Babbush et al. have exponentially accelerated the prediction of oscillator dynamics by Hamiltonian simulations of vast numbers of interconnected harmonic oscillators \cite{babbush2023exponential}. Several methods have also been devised to unitarize and compute non-unitary linear differential equation systems. Javier et al. have solved the Black-Scholes equation by constructing an extended system of non-unitary systems to be unitary \cite{gonzalez2023efficient}. Shi et al. have solved the diffusion equation by adding a new dimension to the spatial dimension of the system and transforming it into a unitary time evolution equation \cite{jin2022quantum,jin2023quantum}. Childs and Wiebe have proposed an algorithm for nonunitary time evolution by stochastically applying unitary linear sums to quantum states \cite{childs2012hamiltonian,an2023quantum}, and a modified form of the method has been shown to achieve a certain kind of theoretical optimum \cite{an2023linear}. These studies demonstrate that all linear differential equation systems can be Hamiltonian simulated on quantum computers. There have also been many reports on how to efficiently implement Hamiltonian simulations \cite{kieferova2019simulating,berry2015simulating,low2017optimal,balasubramanian2020quantum,mcardle2019variational,zhang2023low,low2019hamiltonian} and how to simulate scientific computations \cite{kharazi2024explicit,jin2024analog,lee2023size} in quantum computers.

The remarkable developments in linear differential equation systems are since the physics underlying quantum computers is itself fundamentally linear. Approaches to nonlinear differential equation systems are still in the primitive stage of research, because these require methods to mathematically transform nonlinear systems into linear systems. Various algorithm ideas have been proposed, including those based on unrealized technologies and unproven prospects \cite{lloyd2020quantum}. Here we focus on algorithms that can be realized with currently available quantum computers. Joseph have proposed an algorithm that uses the Koopman operator to represent a finite-dimensional nonlinear system by an infinite linear system and then truncates part of the operator \cite{joseph2020koopman}. This method is arbitrary in the truncation of the operator and in the choice of basis functions that are well matched to the characteristics of the nonlinear system. Even though the choice of basis functions is important not only for the Koopman operator but also for the efficiency of Hamiltonian simulations \cite{shi2024koopman}, the protocol is not clearly stated. 

Using Carleman linearization, Liu et al. have proposed an algorithm that transforms a nonlinear differential equation into an infinite array of linear differential equations and then truncates part of the array \cite{liu2021efficient}. Although this method is arbitrary in the truncation of the array, the mathematical background of the basis functions is clear. Furthermore, they have quantified the nonlinearity by introducing a parameter that represents the ratio between the tendency of the system toward chaos and the friction that keeps the system on orbit and have proven that the equations can be solved within a certain range of parameters. However, algorithms based on Carleman linearization have been pointed out to have evolution time limitations. Sanavio et al. have employed the Carleman linearization for each of the lattice Boltzmann method, Navier-Stokes equations, and Grad formulations, and have performed classical fluid simulations \cite{sanavio2024three}. In all simulations, the solutions diverge when a certain evolution time is reached. Other similar reports also indicate the limit to the evolution time \cite{itani2024quantum,itani2022analysis}. Theoretical analysis of this limit suggests that divergence is an unavoidable numerical phenomenon destined for Hamiltonian simulations applying Carleman linearization, casting doubt on the practicality of the algorithm.

We propose divergence-free simulation methods for nonlinear differential equations in quantum computers. First, for a state $\bm{x}$ to be solved, we assume a pivot state $\bm{s}$ in its neighborhood. Since $\bm{x}$ is expected to behave linearly in the neighborhood of $\bm{s}$, Hamiltonian simulations are possible. Next, each time $\bm{x}$ leaves the neighborhood of $\bm{s}$, $\bm{s}$ is moved to the neighborhood of $\bm{x}$. By repeating the above, the divergence-free algorithm is achieved. Although quantum state tomography measurements of $\bm{x}$ are necessary to update $\bm{s}$, the cost of measurements can be suppressed to a certain level because $\bm{s}$ only needs to satisfy the condition that it is in the neighborhood of $\bm{x}$. Furthermore, Carleman linearization after approximating the behavior of $\bm{x}$ in the neighborhood of $\bm{s}$ with nonlinear polynomials can effectively reduce the number of updates of $\bm{s}$, which is directly related to the cost. As shown below, the proposed methods are able to track the behavior of the solutions of nonlinear equations on quantum computers over a long period of time.

\section{Problem Statement}
We focus on the evolution time limit that comes to quantum algorithms based on Carleman linearization (see Appendix for details of the algorithm). We clarify the problem by applying the algorithm to a simple nonlinear differential equation. The logistic equation is 
\begin{align}
\frac{dx}{dt}=x(1-x),
\label{eq:logi}
\end{align}
where $t$ is time. Carleman linearization of Eq. (\ref{eq:logi}) yields the following infinite-dimensional linear differential equation
\begin{align}
\frac{dy_k}{dt}=k(y_k-y_{k+1}) (k>0),
\end{align}
where $y_k$ is a real number satisfying $y_k(t=0)=x(t=0)^k$.
The infinite-dimensional linear differential equation is approximated in $K+1$ dimensions by setting $k$ to a finite value less than or equal to $K$ and $y_k (t)=0$ for $K>k$.
For $K=3$, the approximate equation is 
\begin{align}
\frac{d}{dt} \begin{pmatrix}y_0\\y_1\\y_2\\y_3\end{pmatrix} = 
\begin{pmatrix}0&0&0&0\\0&1&-1&0\\0&0&2&-2\\0&0&0&3\end{pmatrix}
\begin{pmatrix}y_0\\y_1\\y_2\\y_3\end{pmatrix}.
\label{eq:logi_carle}
\end{align}
Figure 1(a) shows the time evolution of $x$ obtained by solving the approximate equation numerically. The analytical solution is plotted as a dotted line. The results from the algorithm based on the Carleman linearization diverge around $t=2.5$, regardless of $K$. This divergence characterizes the evolution time limit. In fact, since the logistic equation is simple, the $K+1$-dimensional approximate equation by Carleman linearization has an analytical solution \cite{sanavio2024three}. The analytical solution shows that $x$ diverges in time evolution above the evolution time limit $t_{\mathrm {lim}} \sim 2.39$, which is consistent with Fig. 1(a). As discussed below, similar divergence also occurs in Hamiltonian simulations of the KPP-Fisher equation and the phase-field method.

Here we state that the evolution time limit is an unavoidable numerical phenomenon destined to conventional algorithms based on Carleman linearization. Our goal is to overcome the evolution time limit by developing divergence-free quantum algorithms.

\begin{figure*}[t]
\includegraphics[width=1\textwidth]{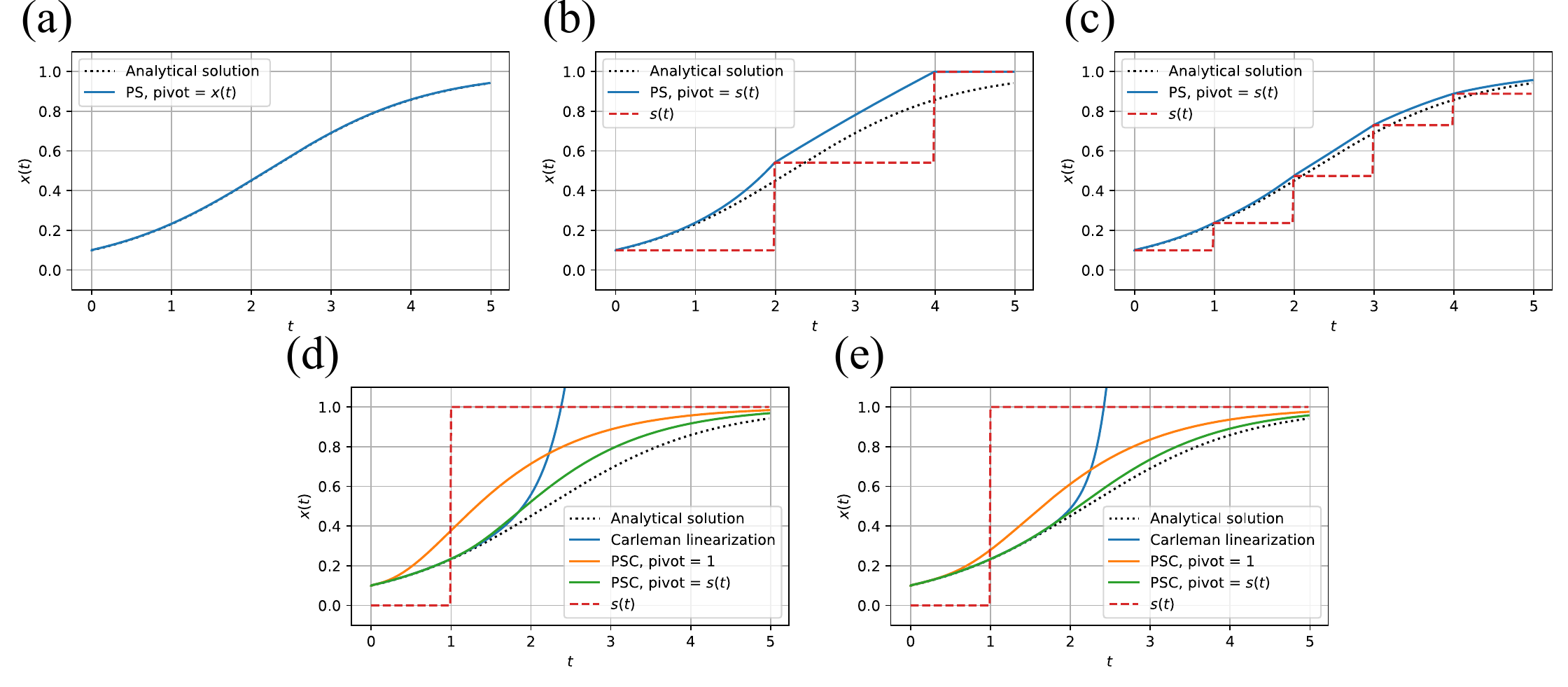}
\caption{
Time evolution of the solution of the logistic equation. (a) Solution of the PS method with $s$ always synchronized with $x$. The dashed line is the analytical solution. (b and c) Solutions of the PS method with $s$ switched every 1 and 2 time intervals, respectively. The red dashed lines are the time variations of $s$. (d and f) Solution of the PSC method with $P$ set to 3 and 5, respectively. The oranges lines are $s$ fixed at 1. The green lines are the initial value of $s$ set to 0 and switched to 1 at t=1. The blue lines are the solution of the conventional Carleman linearization-based algorithm.
}
\label{fig:2}
\end{figure*}

\section{Algorithms}
We present the pivot switching (PS) method as a fundamental divergence-free algorithm for solving nonlinear differential equations on quantum computers. The nonlinear time differential equation for a state $\bm{x}$ represented by an $n$-dimensional real vector is
\begin{align}
&\frac{d\bm{x}}{dt}= \sum_{m=0}^{M}{\bm{F}_m\bm{x}^{\otimes m}} \nonumber \\
&=\bm{F}_{0}+\bm{F}_{1}{\bm{x}}+\bm{F}_{2}\bm{x}^{\otimes 2}+\ldots+\bm{F}_M \bm{x}^{\otimes M}
\label{eq:nonlineq}
\end{align}
\begin{align}
\bm{x}^{\otimes m}=\underset{m\ times}{\underbrace{\bm{x}\otimes\cdots\otimes\bm{x}}},
\end{align}
where $t$ is time, $M$ is a natural number, $x^{\otimes m}$ is an $n^m$-dimensional vector, and $\bm{F}_m \in \mathbb{R}^n \times \mathbb{R}^{n^m}$ is a constant matrix (coefficient matrix of $\bm{x}^{\otimes m}$). For convenience, $\bm{x}^{\otimes 0}=1$. From Eq. (\ref{eq:nonlineq}), the time derivative of state $\bm{
x}$ is the $M$-th order polynomial of $\bm{
x}$ unless $\bm{F}_M$ is a zero matrix. Eq. (\ref{eq:nonlineq}) is a nonlinear differential equation, but it can be approximated as a linear differential equation in the neighborhood of a certain state $\bm{x}=\bm{s}$. Here the state $\bm{s}$ is called a pivot state. Based on the approximation, we introduce the tangent plane of the pivot state, i.e., 
\begin{align}
\sum_{m=0}^{M}{\bm{F}_m\bm{x}^{\otimes m}}=\bm{G}_\mathbf{0}+\bm{G}_\mathbf{1}\left(\bm{x}-\bm{s}\right)+{\bm{O}}\left(\left(\bm{x}-\bm{s}\right)^{\otimes2}\right),
\end{align}
where $\bm{G}_0$ and $\bm{G}_1$ are constant matrices, and $O((\bm{x}-\bm{s})^{\otimes a})$ is an asymptotic notation indicating that the error order is about $a$ power of $\bm{x-s}$ ($a$ is an integer). The linearly approximated nonlinear differential equation is 
\begin{align}
\frac{d\bm{x}}{dt}=\bm{G}_0+\bm{G}_1\left(\bm{x}-\bm{s}\right)=(\bm{G}_\mathbf{0}-\bm{G}_1 {\bm{s}})+\bm{G}_1 {\bm{x}}.
\label{eq:linapprox}
\end{align}
Therefore, when state $\bm{x}(t)$ is in the neighborhood of the pivot state, the nonlinear time differential equation can be approximated by performing a Hamiltonian simulation of equation (\ref{eq:linapprox}). When $\bm{
x}$ leaves the neighborhood of the pivot state, equation (\ref{eq:linapprox}) is no longer valid. Therefore, each time $\bm{
x}$ leaves the neighborhood of $\bm{
s}$, $\bm{
s}$ is moved to the neighborhood of $\bm{
x}$, as shown in Figure 1(b). In other words, the pivot state is updated every time interval $T$ so that the approximate solution obtained by Eq. (\ref{eq:linapprox}) does not exceed an allowed error $E$. Updating the pivot state determines the cost of the PS method because it requires the measurements of $\bm{x}$ by quantum state tomography. Importantly, $\bm{
s}$ need only be a neighborhood of $\bm{
x}$. This means that the determination of $\bm{
s}$ does not require an exact measurement of $\bm{
x}$, and the cost is kept to a certain level. This switching of pivot states updates the valid range of Eq. (\ref{eq:linapprox}) and allows the Hamiltonian simulation to continue. Through repeated switching, a divergence-free algorithm is achieved. From Eq. (\ref{eq:linapprox}), the linear differential equation to be solved by the PS method is
\begin{align}
\frac{d}{dt}\left(\begin{matrix}1\\\bm{x}\\\end{matrix}\right)=\left(\begin{matrix}0&\mathbf{0}\\\bm{G}_\mathbf{0}-\bm{G}_1\bm{s}&\bm{G}_1\\\end{matrix}\right)\left(\begin{matrix}1\\\bm{x}\\\end{matrix}\right),
\end{align}
for the state $\bm{y}_{\mathrm{PS}}=(1, \bm{x})$. The effectiveness of the PS method is discussed below.

The number of times the pivot state $\bm{s}$ is switched in the PS method is directly related to the number of measurements of state $\bm{x}$ and should be reduced. We present the pivot switching Carleman (PSC) method as an advanced algorithm to increase the efficiency of the PS method. While the PS method approximates the $\bm{x}=\bm{s}$ neighborhood linearly with a tangent plane, the PSC method approximates the $\bm{x}=\bm{s}$ neighborhood with a polynomial surface, i.e., 
\begin{align}
\sum_{m=0}^{M}{\bm{F}_m\bm{x}^{\otimes m}}=\sum_{p=0}^{P}{\bm{G}_p^1{(\bm{x}-\bm{s})}^{\otimes p}}+{\bm{O}}\left(\left(\bm{x}-\bm{s}\right)^{\otimes P+1}\right),
\label{eq:nonlinapprox_oterm}
\end{align}
where $\bm{G}_p^1 \in \mathbb{R}^n\times\mathbb{R}^{n^p}$ and $P$ is the degree of the polynomial approximation.
The approximated nonlinear differential equation is 
\begin{align}
\frac{d\bm{x}}{dt}=\sum_{p=0}^{P}{\bm{G}_p^1{(\bm{x}-\bm{s})}^{\otimes p}}.
\label{eq:nonlinapprox}
\end{align}

Although Eq. (\ref{eq:nonlinapprox}) is expected to have higher approximation accuracy than the linear approximation of the PS method, Hamiltonian simulations cannot be performed because it is a nonlinear differential equation. Therefore, we consider the time evolution of the state $\bm{y}_k=x^{\otimes k}$ for all nonnegative integers $k$ as in the Carleman linearization. The time derivative of state $\bm{y}_k$ is also approximated by a polynomial surface, i.e.,
\begin{align}
&\frac{d\bm{y}_k}{dt}=\frac{d\bm{x}^{\otimes k}}{d\bm{x}}\sum_{m=0}^{M}{\bm{F}_m\bm{x}^{\otimes m}}\nonumber \\
&=\sum_{p=0}^{P}{\bm{G}_p^k{(\bm{x}-\bm{s})}^{\otimes p}}+{\bm{O}}\left(\left(\bm{x}-\bm{s}\right)^{\otimes P+1}\right),
\end{align}
where $\bm{G}_p^k\in\mathbb{R}^{n^k}\times\mathbb{R}^{n^p}$.
The approximated equation is 
\begin{align}
\frac{d\bm{y}_k}{dt}=\sum_{p=0}^{P}{\bm{G}_p^k{(\bm{x}-\bm{s})}^{\otimes p}}.
\label{eq:nonlinapprox_y}
\end{align}

The right-hand side is obviously a linear mapping of the vector $(1,\bm{y}_1,\cdots,\bm{y}_P)$, so Eq. (\ref{eq:nonlinapprox_y}) is a linear differential equation for the state $\bm{y}_{\mathrm{PSC}}=(1,\bm{y}_1,\cdots,\bm{y}_P)$. By performing the Hamiltonian simulation of Eq. (\ref{eq:nonlinapprox_y}), the nonlinear time differential equation in Eq. (\ref{eq:nonlinapprox}) can be approximately solved. It is not necessary to calculate the time evolution of $\bm{y}_k$ with $k > P$, since $\bm{y}_k$ with $k>P$ does not contribute to the time evolution of $\bm{y}_1$. Even with polynomial surfaces, Eqs. (\ref{eq:nonlinapprox}) and (\ref{eq:nonlinapprox_y}) are no longer valid when $\bm{x}$ moves away from the neighborhood of the pivot state. Therefore, as with the PS method, each time $\bm{x}$ leaves the neighborhood of $\bm{s}$, $\bm{s}$ is moved to the neighborhood of $\bm{x}$ (see Figure 1(c)). Since polynomial surfaces yield higher approximation accuracy than planar surfaces, the PSC method is expected to effectively reduce the number of times $\bm{s}$ is determined. By a small number of repetitions of switching, a divergence-free and efficient algorithm is achieved. The effectiveness of the PSC method is discussed below.

\begin{figure*}[t]
\includegraphics[width=1\textwidth]{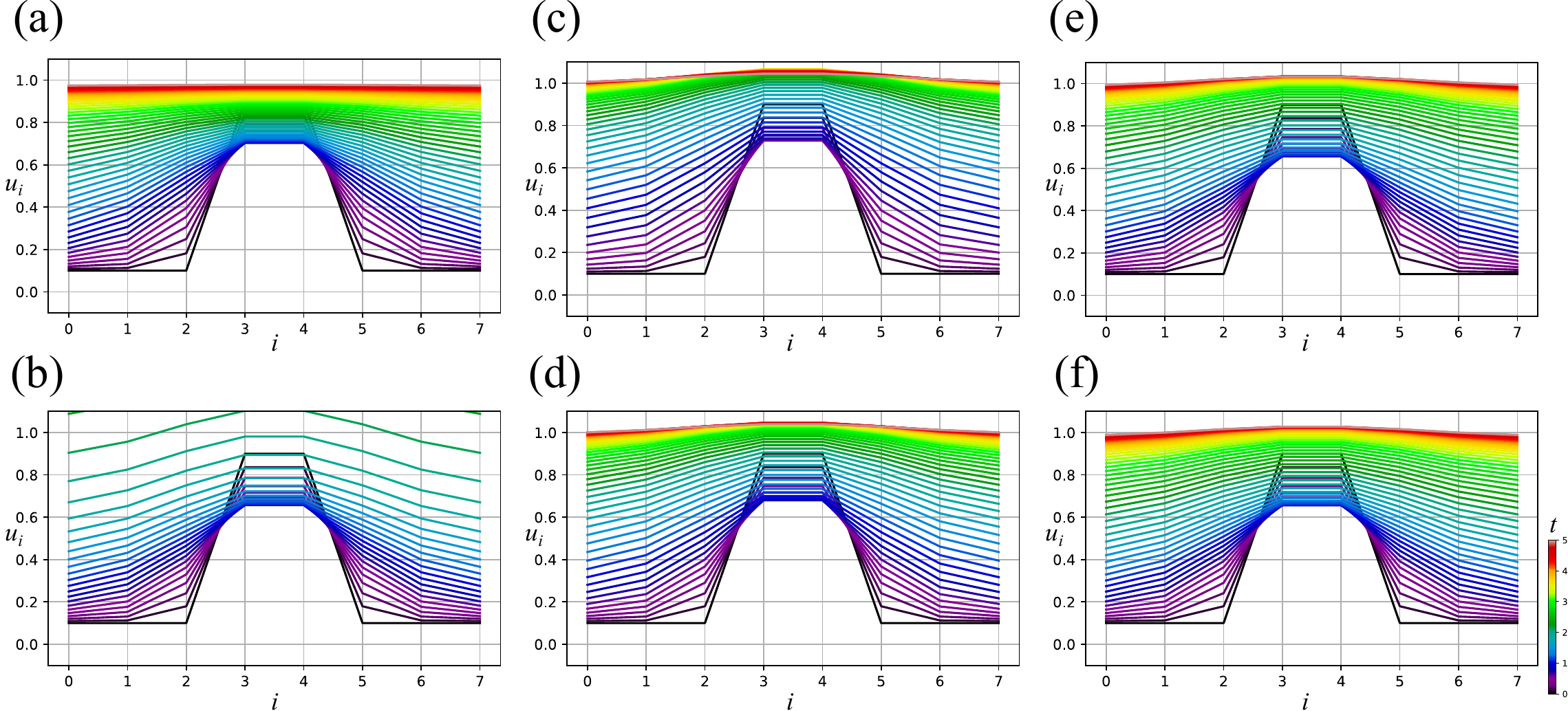}
\caption{
Time evolution of the solution of the KPP-Fisher equation. (a) Classical computational solution. (b) Solution of the conventional Carleman linearization-based algorithm with $K=3$. (c and d) Solution of the PSC method with $P$ set to 3 and 5, respectively, and $\bm{s}$ fixed to $\bm{1}$. (e and f) Solution of the PSC method with $P$ set to 3 and 5, respectively, and $S$ initially set to $\bm{\phi}(t=0)$ and switched to $1$ at $t=1$.}
\label{fig:3}
\end{figure*}

\section{Applications}
Our algorithms can break through the evolution time limit imposed on the conventional algorithm based on Carleman linearization. To demonstrate this, our algorithms were applied not only to the logistic equation illustrated in Problem Statement, but also to the reaction-diffusion equation found in ecology, physiology, combustion and plasma physics, and to the phase-field model applied to general phase-transition problems. The linear equations obtained by our algorithms were integrated by the first-order forward Euler method with time step size $\Delta t=0.01$ in all the applications below.

\subsection{Logistic Equation}
As mentioned in Problem Statement, the conventional Carleman linearization-based algorithm cannot solve the logistic equation over long evolution times in Hamiltonian simulations, so the PS and PSC methods were applied. With the PS method, the logistic equation was approximated as follows, 
\begin{align}
\frac{dx}{dt}=s^2-\left(1-2s\right)x,
\end{align}
where $s$ is the pivot state. The ideal computational accuracy of the PS method is achieved by always having $s(t)=x(t)$. Figure 2(a) shows the result of the PS method when $s(t)=x(t)$ always holds. The ideal PS method is in perfect agreement with the analytical solution (cf. Figure 1(a)). However, from a practical point of view, obtaining the state $x(t)$ from the quantum state is not easy. This is because it requires estimation of $x(t)$ by quantum state tomography or other methods, and multiple independent Hamiltonian simulations must be repeated many times. Moreover, while time-independent Hamiltonian simulations can be implemented efficiently \cite{low2019hamiltonian}, efficient implementations of time-dependent Hamiltonian simulations are not known. For these reasons, the number of times to switch pivot states should be reduced as much as possible. Figures 2(b) and 2(c) show the results of the PS method with the pivot state switched every 1 and 2 time intervals, respectively. The time-dependent pivot state $s(t)$ is also plotted. Switching at every 1 time interval produced results close to the analytical solution (Fig. 2(b)), while switching at every 2 time interval showed a slight deviation from the analytical solution (Fig. 2(c)). Note that this deviation is acceptable depending on the degree of tolerance $E$. It is worth noting that in both cases, the evolution time limit of the conventional Carleman linearization-based algorithm was clearly surpassed.

By the PSC method, the logistic equation was approximated as follows, 
\begin{align}
\frac{dy_k}{dt}=\sum_{p=0}^{P}{g_p^k\left(x-s\right)^p}\ 
\label{eq_logi_psc1}
\end{align}
\begin{align}
g_p^k=\frac{1}{p!}\left. \frac{d^p\left(x\left(1-x\right)\ \cdot k x^{k-1}\right)}{dx^p}\right|_{x=s}.
\label{eq_logi_psc2}
\end{align}
From Eqs. (\ref{eq_logi_psc1}) and (\ref{eq_logi_psc2}), the linear differential equation to be solved was obtained. For example, when $P=3$, the logistic equation is approximated as follows,
\begin{align}
\frac{d}{dt} \begin{pmatrix}y_0\\y_1\\y_2\\y_3\end{pmatrix} = 
\begin{pmatrix}0&0&0&0\\0&1&-1&0\\0&0&2&-2\\3 s^4&-12 s^3&18s^2&3-12s\end{pmatrix}
\begin{pmatrix}y_0\\y_1\\y_2\\y_3\end{pmatrix}.
\label{eq:logi_psc}
\end{align}
Figures 2(d) and (e) show the results of the PSC method when $P=3$ and 5, respectively. The analytical solution, the conventional Carleman linearization, and the time-dependent pivot state $s(t)$ are also plotted. The pivot state of the PSC method was either fixed at 1 or switched from 0 to 1 once at $t=1$. Surprisingly, the PSC method broke the evolution time limit without switching the pivot state, and the obtained solution was acceptable depending on the tolerance $E$. Furthermore, switching the pivot state only once yielded results closer to the analytical solution. Increasing the degree of the polynomial approximation $P$ also improved the accuracy of the results. In particular, when $P=5$ and the pivot state was switched once, the accuracy was equivalent to that of the PS method which switched the pivot state every 1 time interval. This means that the PSC method reduced the number of times the pivot state was switched to 1/5 without any loss of accuracy. Based on this finding, the following sections focus on the evaluation of the PSC method.

\begin{figure*}[t]
\includegraphics[width=1\textwidth]{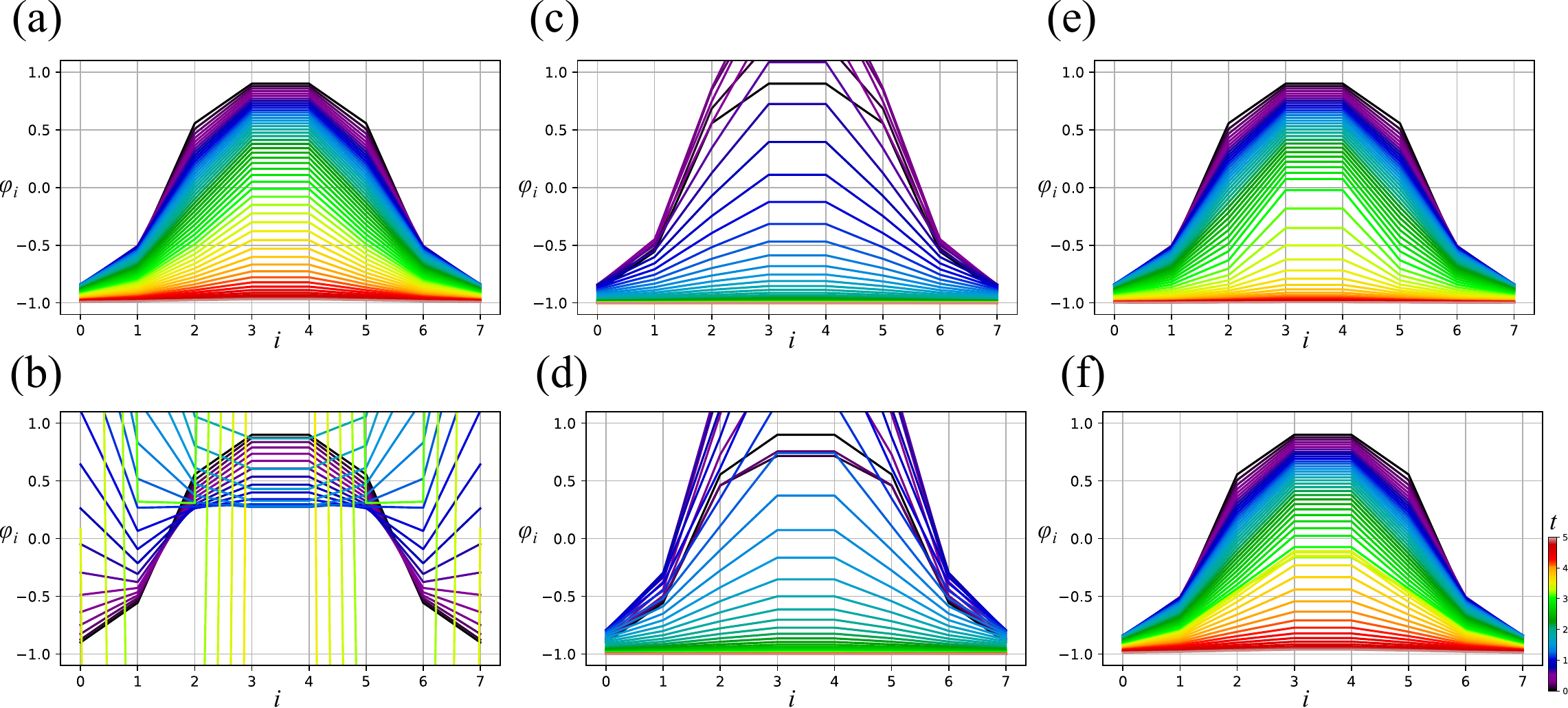}
\caption{
Time evolution of the solution of the phase field model. (a) Classical computational solution. (b) Solution of the conventional Carleman linearization-based algorithm with $K=3$. (c and d) Solution of the PSC method with $P$ set to 3 and 5, respectively, and $\bm{s}$ fixed to $\bm{1}$. (e and f) Solution of the PSC method with $P$ set to 3 and 5, respectively, and $\bm{s}$ initially set to $\bm{\phi}(t=0)$ and switched to $\bm{1}$ at $t=2.9$.}
\label{fig:4}
\end{figure*}

\subsection{Reaction-Diffusion Equation}
The KPP-Fisher equation is a class of reaction-diffusion equations that can exhibit traveling wave solutions that switch between equilibrium states and has been applied to problems in ecology and physiology, combustion and plasma physics, and crystallization. The KPP-Fisher equation is expressed as follows,
\begin{align}
\frac{\partial u(z,t)}{\partial t}=\frac{\partial^2u}{\partial z^2}+u\left(1-u\right),
\label{eq:kpp}
\end{align}
where $u(z,t) \in\mathbb{R}$ is the traveling wave solutions. The first term on the right-hand side is a linear diffusion term and the second term on the right-hand side is a second-order nonlinear source term. Eq. (\ref{eq:kpp}) is a differential equation with spatial coordinate $z$ as a continuous variable, which was difficult to compute directly, so it was discretized with respect to $z$. 
Assuming that the computational point $i \in \mathbb{Z}$, equally spaced with width $\Delta z$, was at coordinate $z=i \Delta z$, and applying second-order precision central differencing to the diffusion term, Eq. (\ref{eq:kpp}) was approximated by the following difference equation,
\begin{align}
\frac{du_i}{dt}=\frac{1}{\Delta z^2}(u_{i-1}-2u_i+u_{i+1})+u_i(1-u_i),
\label{eq:kpp_di}
\end{align}
where $u_i \in \mathbb{R}$. By setting $\Delta z^2=1$ and imposing the periodic boundary condition $u_i=u_{i+8}$, Eq. (\ref{eq:kpp_di}) became the finite-dimensional nonlinear differential equation for state $\bm{u} = (u_1,u_2,\cdots,u_7) \in \mathbb{R}^8$. This was solved using the Carleman linearization and the PSC method and compared with the classical numerical solution.
The initial condition for the simulation was
\begin{align}
u_i(t=0)=\ \left\{\begin{matrix}0.9&i=3,4\\0.1&{\rm otherwize}\\\end{matrix}\right. .
\end{align}

Figure 3(a) is the classical numerical solution, which is considered the most accurate result. Figure 3(b) is the result from Carleman linearization for $K=3$. Similar to the logistic equation results (Fig. 1(a)), the algorithm based on the conventional Carleman linearization diverged when the evolution time limit was reached. Figures 3(c) and 3(d) show the results of the PSC method with $P=3$ and 5, respectively, and with the pivot state fixed at 1. For the KPP-Fisher equation, the PSC method also broke the evolution time limit without switching the pivot state, and the obtained solution was acceptable depending on the tolerance $E$. Figures 3(e) and 3(f) show the results of the PSC method with $P=3$ and 5, respectively, with the pivot state set to the same value as the initial condition and switched to 1 only once at $t=1$. By switching the pivot state only once, the solution approached that of the classical numerical method. In particular, when $P=5$ and the pivot state was switched only once, $u(t)$ was in good agreement with those of the classical numerical simulation. This means that the KPP-Fisher equation can be solved on quantum computers by the PSC method without divergence or loss of accuracy.

\subsection{Phase-Field Model}
The phase field model is one of the practical models applied to general phase transition problems. We considered the model expressed as follows,
\begin{align}
\frac{d\phi(z,t)}{dt}=\frac{d^2\phi}{dz^2}+\left(\phi-1\right)\left(\phi+\beta\right)\left(\phi+1\right),
\label{eq:pf}
\end{align}
where $\phi(z,t)$ is the order parameter. The first term on the right-hand side is a linear term and the second term on the right-hand side is a third-order nonlinear reaction term. The reaction term drives the order parameter to 1 or -1. $\beta$ determines the stability of the order parameter: if $\beta > 0$, $\phi=1$ is stable; if $\beta < 0$, $\phi=-1$ is stable.
Eq. (\ref{eq:pf}) is a differential equation with spatial coordinate $z$ as a continuous variable, so we discretized it with respect to $z$. Assuming that the computational point $i$, equally spaced with width $\Delta z$, was at coordinate $z=i\Delta z$, and applying second-order precision central differencing to the diffusion term, Eq. (\ref{eq:pf}) was approximated by the following difference equation,
\begin{align}
\frac{d\phi_i}{dt}= &
\frac{1}{\Delta z^2}(\phi_{i-1}-2\phi_i+\phi_{i+1}) + \nonumber \\ 
& \left(\phi_i-1\right)\left(\phi_i+\beta\right)\left(\phi_i+1\right)\ 
\label{eq:pfdi}
\end{align}
By setting $\Delta z^2=1$ and imposing the periodic boundary condition $\phi_i=\phi_{i+8}$, equation (\ref{eq:pfdi}) became the finite-dimensional nonlinear differential equation for the state $\bm{\phi}=(\phi_0,\phi_1,\cdots,\phi_7 )\in \mathbb{R}^8$. This equation was solved using the Carleman linearization and PSC methods and compared with the classical numerical solution. $\beta$ was set to $-0.2$ and the initial condition for the simulation was
\begin{align}
& {\bm{\phi}}\left(t=0\right)= \nonumber \\ & \left(-0.90,\ -0.56,\ 0.56,\ 0.90,\ 0.90,\ 0.56,\ -0.56,\ -0.90\right). \nonumber \\ \
\end{align}
This initial condition was set so that $\bm{\phi}(t) = \bm{\phi}(0)$ when $\beta=0$, i.e., an equilibrium state.

Figure 4(a) is the classical numerical solution, which is considered the most accurate result. Since $\beta<0$, the solution evolved to $\bm{\phi}=-\bm{1}$. Figure 4(b) is the result from the Carleman linearization for $K=3$. Similar to the results for the logistic equation (Fig. 1(a)) and the KPP-Fisher equation (Fig. 3(b)), the algorithm based on the conventional Carleman linearization diverged when the evolution time limit was reached. Fig. 4(c) and (d) show the results of the PSC method with $P=3$ and 5, respectively, and with the pivot state fixed at -1. In the phase-field model, the PSC method broke the evolution time limit without switching the pivot state. However, the solution deviations in the early simulation time were significant; recall that the theoretical range of values for $\phi$ is from -1 to 1. Figures 4(e) and (f) show the results of the PSC method with $P=3$ and 5, respectively, with the pivot state set to the same value as the initial condition and switched to $-\bm{1}$ only once at $t=2.9$. By switching the pivot state only once, the solution approached that of the classical numerical solution. In particular, when $P=5$ and the pivot state was switched only once, the phase fields $\bm{\phi}(t)$ was in good agreement with those of the classical numerical simulation. This means that the phase-field model can be solved on a quantum computer by the PSC method with sufficient accuracy and without divergence.

\section{Discussion}
In this study, we have presented the PS method as the fundamental divergence-free algorithm for solving nonlinear differential equations on quantum computers. Furthermore, we have presented the PSC method as the advanced algorithm to improve the efficiency of the PS method. These algorithms break through the evolution time limit destined for the conventional Carleman linearization-based quantum algorithm, as confirmed by the logistic equation, the KPP-Fisher equation, and the Phase-field model. Applications of the PS method to logistic equations have shown that the basic idea of switching pivot states makes a substantial contribution to achieving divergence-free algorithms. The more frequent the switching, the closer the solution by the PS method is to the analytical solution, but the further away the method is from practicality. This trade-off has been greatly mitigated by the PSC method. The idea of approximating the $\bm{x}=\bm{s}$ neighborhood nonlinearly with polynomial surface rather than linearly with tangent plane and then solving by Carleman linearization has been shown to significantly reduce the number of pivot state updates while maintaining solution accuracy. Applied to the KPP-Fisher equation and phase-field model, the PSC method not only has achieved divergence-free Hamiltonian simulations, but also has yielded solutions that close to those of classical numerical simulations with only one coarse pivot-state switching operation. This means that practical problems described by nonlinear differential equation systems can be solved on quantum computers with sufficient accuracy using the PSC method.

The mathematical differences between the PSC method and the conventional Carleman linearization-based algorithm are interesting. For example, comparing Eqs. (\ref{eq:logi_carle}) and (\ref{eq:logi_psc}), the only difference is in the highest order $y_3$ differential equation. The difference in the results produced by the two methods is due to the behavior of the solution of the highest-order equation. This is also true for general nonlinear systems of equations. When $M<P$, the terms below the $P-M$ order are guaranteed to be below the $P$ order (see Eq. (\ref{eq:nonlinapprox_oterm})), so the polynomial approximation by the PSC method of the differential equations of $y_k \ (k<M-P)$ is invariant regardless of the pivot state. Thus, stabilizing the behavior of higher order terms in a finite Carleman linearization is a prescription to prevent catastrophic results due to solution divergence. The PSC method is one method for stabilizing the behavior and presents the possibility of other methods based on similar concepts.

The PS and PSC methods rely on “loose measurements” to obtain an estimate of state $\bm{x}$ with error from the small number of measurement operations, which is then used as the pivot state $\bm{s}$. This paper has demonstrated that loose measurements are sufficient, since $\bm{s}$ only needs to satisfy the condition that it is in the neighborhood of $\bm{x}$. Although loose measurement can reduce the cost of measurement to the certain level, the PSC method has been proposed for further cost reduction. Thus, loose measurements are a necessary cost to prevent solution divergence. This cost could be further reduced by combining loose measurements with weak measurements \cite{dressel2014colloquium}.

Here let us emphasize other potential advantages of loose measurement, namely its applicability to nonlinear systems that tend more toward chaos. For example, when the chaotic solution moves back and forth between attractors, $\bm{s}$ leaves $\bm{x}$, so the PS and PSC methods measure $\bm{s}$ loosely. If the frequency of $\bm{s}$ updates is appropriate, then the approximation used by our algorithms hold even if the solution is chaotic. This raises the possibility that the PS and PSC methods can be applied to nonlinear systems that tend more toward chaos. A detailed investigation in this regard is the subject of future work.

A possible future strategy is to determine the pivot state directly by quantum computation rather than obtaining it from loose measurements. Under quantum superposition, multiple candidate pivot states are employed, and Hamiltonian simulations are performed in parallel. By using an amplitude amplification algorithm \cite{brassard2002quantum} to remove only those candidate pivot states that move toward divergence, the pivot state can be determined without loose measurements. We expect that our methods, in combination with other promising methods, lead to efficient and robust quantum algorithms for nonlinear systems of equations.

Finally, let us emphasize that the algorithms presented in this paper are significant in that these can track the behavior of the solution to systems of nonlinear equations over the long period of time. For example, in general phase transition problems, the behavior of the solution over time is important for understanding and predicting the phenomenon. This importance will become more pronounced when quantum computers can be applied to larger nonlinear systems, i.e., large systems that represent phenomena in two- or three-dimensional spatial coordinate systems. Furthermore, long-time simulations will become even more important as their suitability for the chaotic behavior described above is demonstrated. We expect that divergence-free quantum algorithms for nonlinear equations represent a milestone in practical applications of quantum computers and will motivate research into more general-purpose algorithms.

\bibliography{apssamp}

\appendix
\section{Conventional Quantum Algorithm based on Carleman Linearization}

The method based on Carleman linearization has been employed to solve Eq. (\ref{eq:nonlineq}) by Hamiltonian simulation. In this method, the time evolution of extended variables $\bm{y}_k = \bm{x}^{\otimes k}$ for all non-negative integer $k$ is taken into account. From Eq. (\ref{eq:nonlineq}), the time evolution equation for $\bm{y}_k$ is
\begin{align}
\frac{d\bm{y}_k}{dt}=\frac{d\bm{x}^{\otimes k}}{dt}=\frac{d\bm{x}^{\otimes k}}{d\bm{x}}\frac{d\bm{x}}{dt}\ =\frac{d\bm{x}^{\otimes k}}{d\bm{x}}\sum_{m=0}^{M}{\bm{F}_m\bm{x}^{\otimes m}}.
\label{eq:apdx_dykdt}
\end{align}
Eq. (\ref{eq:apdx_dykdt}) is further expanded as
\begin{align}
&\frac{d\bm{y}_k}{dt} = \sum_{l=k}^{k+M}{A_{k,l}\bm{y}_l}  (k \geq 0),
\label{eq:apdx_dykdtA}
\end{align}
where $A_{k,l} \in \mathbb{R}^{n^k} \times \mathbb{R}^{n^l}$ is a constant coefficient matrix defined as
\begin{align}
A_{k,l}=\left(\sum_{v=0}^{k-1}\underset{v\ times}{\underbrace{I\otimes\cdots\otimes I}}\otimes\bm{F}_{l-k-1}\otimes\underset{k-1-v\ times}{\underbrace{I\otimes\cdots\otimes I}}\right).
\end{align}
The following relationship is used to derive equation (\ref{eq:apdx_dykdtA}):
\begin{align}
\frac{d\bm{x}^{\otimes k}}{d\bm{x}}=\sum_{v=0}^{k-1}\underset{v\ times}{\underbrace{\bm{x}\otimes\cdots\otimes\bm{x}}}\otimes I \otimes \underset{k-1-v\ times}{\underbrace{\bm{x}\otimes\cdots\otimes\bm{x}}}.
\end{align}

Eq. (\ref{eq:apdx_dykdtA}) is the system of linear equations for infinite-dimensional vector $\bm{y}_\mathrm{inf} = (y_0, \bm{y}_1, \bm{y}_2, \cdots)$. 
Eq. (\ref{eq:apdx_dykdtA}) implies that the solution to the nonlinear differential equation (Eq. (\ref{eq:nonlineq})) can be obtained by solving the system of infinite-dimensional linear differential equations.
However, since $y_\mathrm{inf}$ is infinite-dimensional, it cannot be solved realistically even if a quantum computer is used. Therefore, the higher order terms in Eq. (\ref{eq:apdx_dykdtA}) are truncated as follows,
\begin{align}
\frac{d\bm{y}_k}{dt}=\sum_{l=k}^{K}{A_{k,l}\bm{y}_l} \ (K \geq k \geq 0).
\label{eq:apdx_final_carle}
\end{align}
Eq. (\ref{eq:apdx_final_carle}) is the system of equations to be solved by the quantum algorithm based on Carleman linearization. Note that even if the system of equations is not unitary, a solution can be obtained by applying non-unitary Hamiltonian simulation methods.

\nocite{*}


\end{document}